\documentclass[fleqn,usenatbib]{mnras}

\usepackage{newtxtext,newtxmath}

\usepackage[T1]{fontenc}

\DeclareRobustCommand{\VAN}[3]{#2}
\let\VANthebibliography\thebibliography
\def\thebibliography{\DeclareRobustCommand{\VAN}[3]{##3}\VANthebibliography}

\usepackage{graphicx}	
\usepackage{amsmath}	

\usepackage{bm} 
\usepackage{chemformula} 
\let\ce\ch 
\usepackage{xspace} 
\newcommand{\kms}{\,km\,s$^{-1}$\xspace} 
\newcommand{\K}{\,K\xspace} 
\newcommand{\GHz}{\,GHz\xspace} 
\newcommand{\kcalmol}{\,kcal\,mol$^{-1}$\xspace} 

\title[The \ce{H2CN}/\ce{H2NC} ratio as a $T_\text{kin}$ tracer for the ISM]{\ce{H2CN}/\ce{H2NC} abundance ratio: a new potential temperature tracer for the interstellar medium}

\author[D. San Andrés et al.]{
D. San Andrés,$^{1}$\thanks{E-mail: david.sanandres@cab.inta-csic.es}
L. Colzi,$^{1}$
V. M. Rivilla,$^{1}$
J. García de la Concepción,$^{2,3}$
M. Melosso,$^{4}$
\newauthor
J. Martín-Pintado,$^{1}$
I. Jiménez-Serra,$^{1}$
S. Zeng,$^{5}$
S. Martín,$^{6,7}$
M. A. Requena-Torres$^{8,9}$
\\
$^{1}$Centro de Astrobiología (CAB), INTA-CSIC, Carretera de Ajalvir km 4, Torrejón de Ardoz, 28850, Madrid, Spain\\
$^{2}$Departamento de Química Orgánica e Inorgánica, Facultad de Ciencias, Universidad de Extremadura, E-06006 Badajoz, Spain\\
$^{3}$IACYS-Unidad de Química Verde y Desarrollo Sostenible, Facultad de Ciencias, Universidad de Extremadura, E-06006 Badajoz, Spain\\
$^{4}$Scuola Superiore Meridionale, Largo San Marcellino 10, 80136 Naples, Italy\\
$^{5}$Star and Planet Formation Laboratory, Cluster for Pioneering Research, RIKEN, 2–1 Hirosawa, Wako, Saitama, 351–0198, Japan\\
$^{6}$European Southern Observatory, Alonso de Córdova, 3107, Vitacura, Santiago 763–0355, Chile\\
$^{7}$Joint ALMA Observatory, Alonso de Córdova, 3107, Vitacura, Santiago 763–0355, Chile\\
$^{8}$Department of Astronomy, University of Maryland, College Park, MD 20742–2421, USA\\
$^{9}$Department of Physics, Astronomy and Geosciences, Towson University, Towson, MD 21252, USA\\
}

\date{Accepted 2023 May 4. Received 2023 May 4; in original form 2023 January 20}

\pubyear{2015}

\begin{document}
\label{firstpage}
\pagerange{\pageref{firstpage}-\pageref{lastpage}}
\maketitle

\begin{abstract}
The \ce{H2NC} radical is the high-energy metastable isomer of \ce{H2CN} radical, which has been recently detected for the first time in the interstellar medium towards a handful of cold galactic sources, besides a warm galaxy in front of the PKS 1830-211 quasar. These detections have shown that the \ce{H2CN}/\ce{H2NC} isomeric ratio, likewise the \ce{HCN}/\ce{HNC} ratio, might increase with the kinetic temperature ($T_\text{kin}$), but the shortage of them in warm sources still prevents us from confirming this hypothesis and shedding light on their chemistry. 
In this work, we present the first detection of \ce{H2CN} and \ce{H2NC} towards a warm galactic source, the G+0.693-0.027 molecular cloud (with $T_\text{kin} > 70$\K), using IRAM 30m telescope observations. We have detected multiple hyperfine components of the $N_{K_\text{a}K_\text{c}} = 1_{01} - 0_{00}$ and $2_{02} - 1_{01}$ transitions. We derived molecular abundances with respect to \ce{H2} of (6.8$\pm$1.3) $\times 10^{-11}$ for \ce{H2CN} and of (3.1$\pm$0.7) $\times 10^{-11}$ for \ce{H2NC}, and a \ce{H2CN}/\ce{H2NC} abundance ratio of $2.2 \pm 0.5$. 
These detections confirm that the \ce{H2CN}/\ce{H2NC} ratio is $\gtrsim$2 for sources with $T_\text{kin} > 70$\K, larger than the $\sim$1 ratios previously found in colder cores ($T_\text{kin}\sim10$\K). 
This isomeric ratio dependence with temperature cannot be fully explained with the currently proposed gas-phase formation and destruction pathways. Grain surface reactions, including the $\ce{H2NC} \rightarrow \ce{H2CN}$ isomerization, deserve consideration to explain the higher isomeric ratios and \ce{H2CN} abundances observed in warm sources, where the molecules can be desorbed into the gas phase through thermal and/or shock-induced mechanisms.
\end{abstract}

\begin{keywords}
ISM: molecules -- ISM: clouds -- astrochemistry -- Galaxy: centre -- line: identification
\end{keywords}



\section{Introduction}\label{sec:introduction}

Since the first detection of \ce{CH} in space by \citet{Swings&Rosenfeld1937} and \citet{McKellar1940}, the rapid development of new instrumentation (especially at infrared and mainly at radio wavelengths) has led up to date to the confirmed detection of about 290 molecules\footnote{\url{https://cdms.astro.uni-koeln.de/classic/molecules}} in the interstellar medium (ISM) or circumstellar shells within our own Galaxy \citep{McGuire2021}. Recently, it has been reported the first detection in space of \ce{H2NC} (aminomethylidyne), the high-energy metastable isomer of the well-known \ce{H2CN} (methylene amidogen radical). 
\citet{Cabezas2021} and \citet{Agundez2023} have detected both isomers towards seven cold dense clouds, and also towards a $z = 0.89$ galaxy in front of PKS 1830-211 quasar (hereafter PKS 1830-211). These authors found that the \ce{H2CN}/\ce{H2NC} abundance ratio is $\sim$1 towards the cold clouds (with gas kinetic temperatures, $T_\text{kin}$, of $\sim$10\K), but higher ($\sim$4) towards the warmer PKS 1830-211 ($T_\text{kin} \gtrsim 80$\K, \citealt{Henkel2008}). This suggests that the \ce{H2CN}/\ce{H2NC} ratio might be sensitive to the kinetic temperature, similarly to the  better studied \ce{HCN}/\ce{HNC} ratio (e.g., \citealt{Irvine&Schloerb1984}; \citealt{Schilke1992}; \citealt{Hacar2020}). The \ce{HCN}/\ce{HNC} abundance ratio has been found to be $\sim$1 towards cold clouds ($T_\text{kin} < 20$\K, \citealt{Sarrasin2010}; \citealt{Colzi2018}), but $\gtrsim$1 in warmer sources (\citealt{Liszt&Lucas2001}; \citealt{Colzi2018}). \citet{Hacar2020} explored the use of the \ce{HCN}/\ce{HNC} abundance ratio as a potential chemical thermometer in a large scale map of the Orion molecular cloud, and found a linear dependence between the \ce{HCN}/\ce{HNC} ratio and the gas $T_\text{kin}$. They found the lowest ratios ($\sim$1$-$4) towards the coldest regions ($T_\text{kin} < 30$\K), while the higher ratios (>5) were associated to the warmer sources ($T_\text{kin} \gtrsim 30$\K). 

\citet{Agundez2023} searched for \ce{H2CN} and \ce{H2NC} towards two warm galactic diffuse clouds with $T_\text{kin} > 30$\K (\citealt{Liszt2006, Liszt2010}; \citealt{Chantzos2020}) unsuccessfully. The lack of detections of \ce{H2NC} and \ce{H2CN} towards warm galactic sources, besides the extragalactic ISM, prevent the confirmation to what extent the \ce{H2CN}/\ce{H2NC} ratio is indeed sensitive to the temperature. New detections of both isomers towards warm sources are required to gain insight about the chemistry of these molecular species.

We present the first detection of both \ce{H2NC} and \ce{H2CN} isomers towards a warm galactic molecular cloud: G+0.693-0.027 (hereafter G+0.693). This source, located within the Galactic Centre Sgr B2 complex, exhibits high $T_\text{kin}$ ($\sim$70$-$140\K, \citealt{Zeng2018}). G+0.693 is suggested to be affected by low-velocity shocks as a consequence of large scale cloud-cloud collisions, responsible for the release into the gas-phase of molecules formed on the dust grain surfaces (\citealt{Martin2008}; \citealt{Zeng2020}).
This produces a extremely rich chemical reservoir; in fact, more than 120 molecular species have been already identified, including many nitrogen-bearing species (\citealt{Requena-Torres2006, Requena-Torres2008}; \citealt{Zeng2018, Zeng2021}; \citealt{Rivilla2018, Rivilla2019, Rivilla2020, Rivilla2021a, Rivilla2021b, Rivilla2022a, Rivilla2022c, Rivilla2022b}; \citealt{Jimenez-Serra2020, Jimenez-Serra2022}; \citealt{Rodriguez-Almeida2021b, Rodriguez-Almeida2021a}; \citealt{Colzi2022}).

This article is organized as follows: in Sect.~\ref{sec:observations} we present the data of the observational survey towards G+0.693, while in Sect.~\ref{sec:analysis and results}
we describe the identification of \ce{H2CN} and \ce{H2NC} and analysis procedures we have followed as well as the summary of the main findings. In Sect.~\ref{sec:discussion} we study the possible link between the \ce{H2CN}/\ce{H2NC} ratio and $T_\text{kin}$, and the potential use of this ratio as a thermometer. We also present a discussion on the interstellar chemistry of \ce{H2CN} and \ce{H2NC}, exploring their main formation and destruction routes. Finally, the conclusions are summarised in Sect.~\ref{sec:conclusions}.

\section{Observations}\label{sec:observations}

We have analysed data from a high sensitivity spectral survey carried out towards G+0.693 (\citealt{Zeng2020}; \citealt{Rivilla2020, Rivilla2021a}). Single-dish observations centered at $\alpha_\text{J2000} = 17^\text{h}47^\text{m}22^\text{s}$ and $\delta_\text{J2000} = -28^\text{o}21'27''$ were performed by using IRAM 30m radiotelescope (Granada, Spain). The observations were gathered in three observational campaigns during 2019 (April 10-16, August 13-19 and December 11-15) and were part of the projects 172-18 (PI Martín-Pintado), 018-19 and 133-19 (PI Rivilla), respectively.  Observations were conducted using position switching mode, with the off position located at an offset of $\Delta\alpha = -885''$, $\Delta\delta = 290''$. 
The broadband heterodyne Eight MIxer Receiver (EMIR) and the Fast Fourier Transform Spectrometer FTS200 were used, which provided a raw frequency resolution of $\sim$200$\, \text{kHz}$. In this work we used data covering the spectral ranges 71.8$-$116.7\GHz and 124.8$-$175.5\GHz. We smoothed the spectra achieving a final resolution of $\sim$800$\, \text{kHz}$ (1.4$-$3.3\kms), which is sufficient to resolve G+0.693 molecular line profiles that present typical line widths of $\sim$15$-$25\kms. 
The telescope beam width varies from $\sim$34.2$''$ at $72 \, \text{GHz}$ up to $\sim$14.0$''$ at $176 \, \text{GHz}$. The line intensity of the spectra was measured in antenna temperature, $T_\text{A}^*$, units, as the molecular emission towards G+0.693 is extended over the beam (e.g., \citealt{Zeng2020}). More details of the observations are provided in \citet{Rivilla2021a,Rivilla2021b,Rivilla2022c}.

\section{Analysis and Results}\label{sec:analysis and results}

The identification of \ce{H2CN} and \ce{H2NC} molecular lines and their fitting has been performed using the version from the $3^\text{rd}$ March 2021 of the Spectral Line Identification and Modeling (SLIM) tool within the MADCUBA\footnote{Madrid Data Cube Analysis on ImageJ is a software developed at the Centre of Astrobiology (CAB) in Madrid: \url{https://cab.inta-csic.es/madcuba}} package \citep{MADCUBA}. SLIM generates a synthetic spectrum under the assumption of local thermodynamic equilibrium (LTE) conditions, to be compared with the observed one. Then, an auto-fitting procedure, SLIM-AUTOFIT, provides the best non-linear least-squares LTE fit to the data using the Levenberg-Marquardt algorithm. The free parameters used for the fit procedure are the total column density of the molecule ($N$), the excitation temperature ($T_\text{ex}$), the local standard of rest velocity ($v_\text{LSR}$), and the full width at half maximum ($\text{FWHM}$). 

For the analysis, we have used the Cologne Database for Molecular Spectroscopy (CDMS, \citealt{Endres2016}) spectroscopic entries 028502 (July 2000) for \ce{H2CN}, and 028528 (September 2021) for \ce{H2NC}, which were obtained from \citet{Yamamoto1992} and \citet{Cabezas2021}, respectively. The dipole moment of \ce{H2CN} was obtained from ab initio calculation by \citet{Cowles1991}, while that of \ce{H2NC} comes form a quantum chemical calculation from the work by \citet{Cabezas2021}. To perform the LTE line fitting, we have used the most unblended transitions of each molecule, along with those partially blended with lines from other molecular species previously identified and modelled towards G+0.693.

\subsection{Detection of \ce{H2NC}}\label{sec:H2NC_detection}

\begin{figure*}
    \centering
    \includegraphics[width=\textwidth]{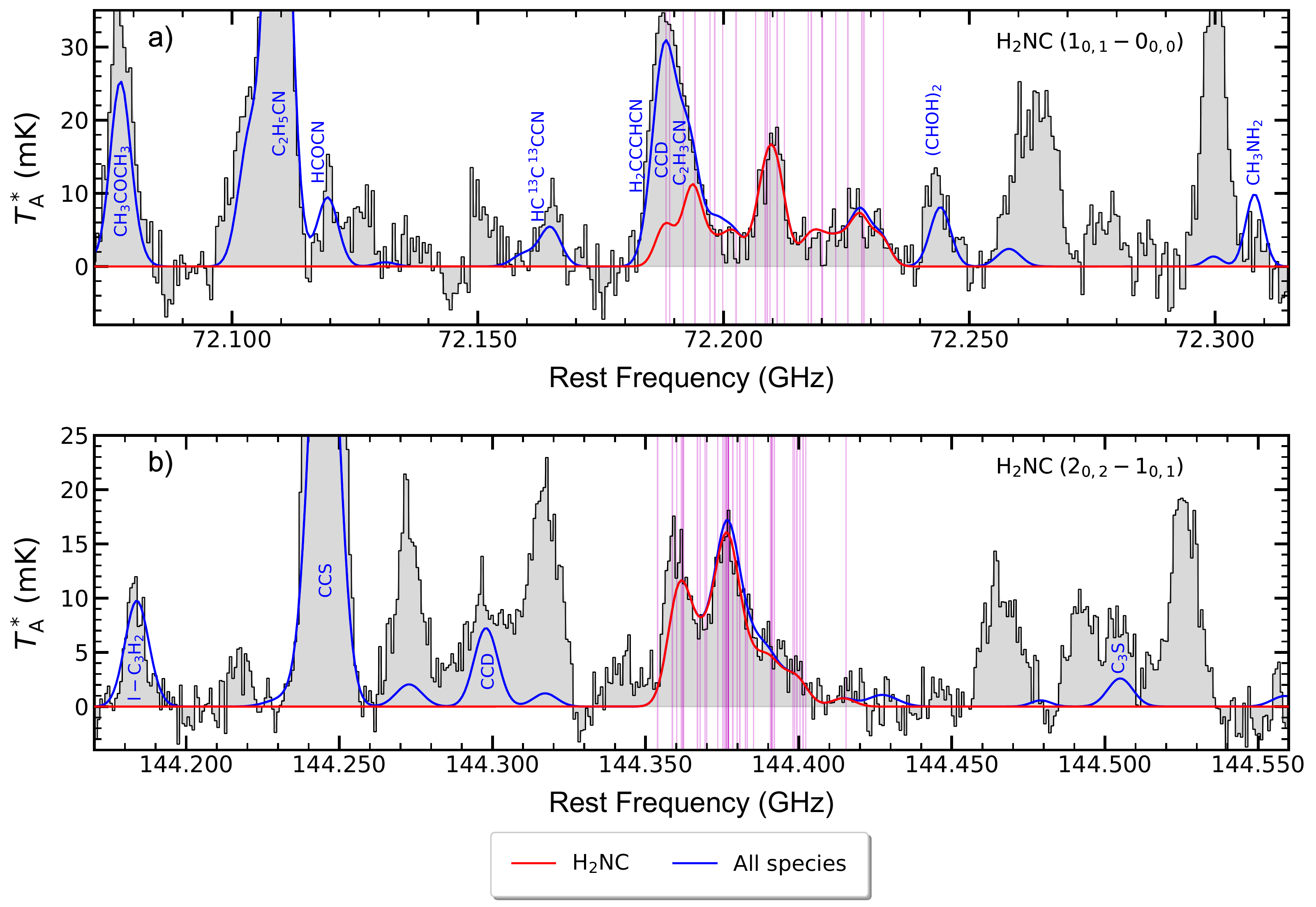}
    \caption{\ce{H2NC} hyperfine lines observed towards G+0.693, corresponding to the $1_{01} - 0_{00}$ rotational transition (upper panel) and the $2_{02} - 1_{01}$ rotational transition (lower panel). Black histogram and grey-shaded areas indicate the observed spectrum, while the red and blue lines represent the best LTE fit obtained for \ce{H2NC}, and the emission of all the species already identified in the cloud (whose names are indicated by the blue labels), respectively. Magenta vertical lines indicate those transitions selected to perform the AUTOFIT. The transitions shown are only those with peak intensities of $T_\text{A}^* > 0.41 \, \text{mK}$, which corresponds to $\sigma/3$ value (where $\sigma$ is the rms of the spectra). The spectroscopic information for the observed transitions is given in Table~\ref{tab:H2NC_spectroscopy} of Appendix~\ref{sec:spectroscopy_tables}.} 
    \label{fig:H2NC_lines} 
\end{figure*}

We show in Fig.~\ref{fig:H2NC_lines} all the \ce{H2NC} transitions detected towards G+0.693 (purple vertical lines of Fig.~\ref{fig:H2NC_lines}), consisting of the hyperfine structure of the $N_{K_\text{a}K_\text{c}} = 1_{01} - 0_{00}$ and $2_{02} - 1_{01}$ rotational transitions, with upper energy levels of 3.5\K and 10.4\K, respectively. Only transitions with intensity $T_\text{A}^* > \sigma/3$ (where $\sigma$ is the rms of the spectra) are considered to compute the synthetic spectrum. The spectroscopic information of the observed transitions is shown in Table~\ref{tab:H2NC_spectroscopy} of Appendix~\ref{sec:spectroscopy_tables}. The pattern produced by all these hyperfine components creates a very characteristic spectral line profile, which reproduces very nicely the observed spectrum, confirming the presence of this molecule in G+0.693. All of the transitions are unblended from emission of other species, except those falling in the 72.19$-$72.20\GHz range, which are blended with \ce{H2CCCHCN}, \ce{CCD} and \ce{C2H3CN} (see Fig.~\ref{fig:H2NC_lines}). The global fit including all species properly matches the observed spectrum.

The LTE AUTOFIT procedure for \ce{H2NC} was carried out by using all its detected transitions and leaving all four parameters free, obtaining $N = (4.2 \pm 0.7) \times 10^{12} \, \text{cm}^{-2}$, $T_\text{ex} = 3.28 \pm 0.11$\K, $v_\text{LSR} = 71.0 \pm 0.5$\kms, and $\text{FWHM} = 18.3 \pm 1.1$\kms (see Table~\ref{tab:fitting_parameters}). As expected, the derived $T_\text{ex}$ is much lower than the $T_\text{kin}$ of the source ($\sim$70$-$140\K; \citealt{Zeng2018}), and similar to that derived for chemically similar species such as the \ce{HCN} and \ce{HNC} isotopologues (\citealt{Colzi2022}), or other species such as \ce{PN} (\citealt{Rivilla2022a}). Higher values of $T_\text{ex}$ (5$-$10\K) do not properly fit both rotational transitions of \ce{H2NC}, but significantly overestimate the $2_{02} - 1_{01}$ transitions. 

We have derived the molecular abundance of \ce{H2NC} with respect to \ce{H2}, using $N_\text{\ce{H2}} = 1.35\times10^{23} \, \text{cm}^{-2}$ from \citet{Martin2008} and assuming an uncertainty of $15\%$ to its value, as done in \citet{Rodriguez-Almeida2021b}. The resulting abundance is (3.1$\pm$0.7) $\times 10^{-11}$ (see Table~\ref{tab:fitting_parameters}).

\subsection{Detection of \ce{H2CN}}\label{sec:H2CN_detection}

\begin{figure*}
    \centering
    \includegraphics[width=\textwidth]{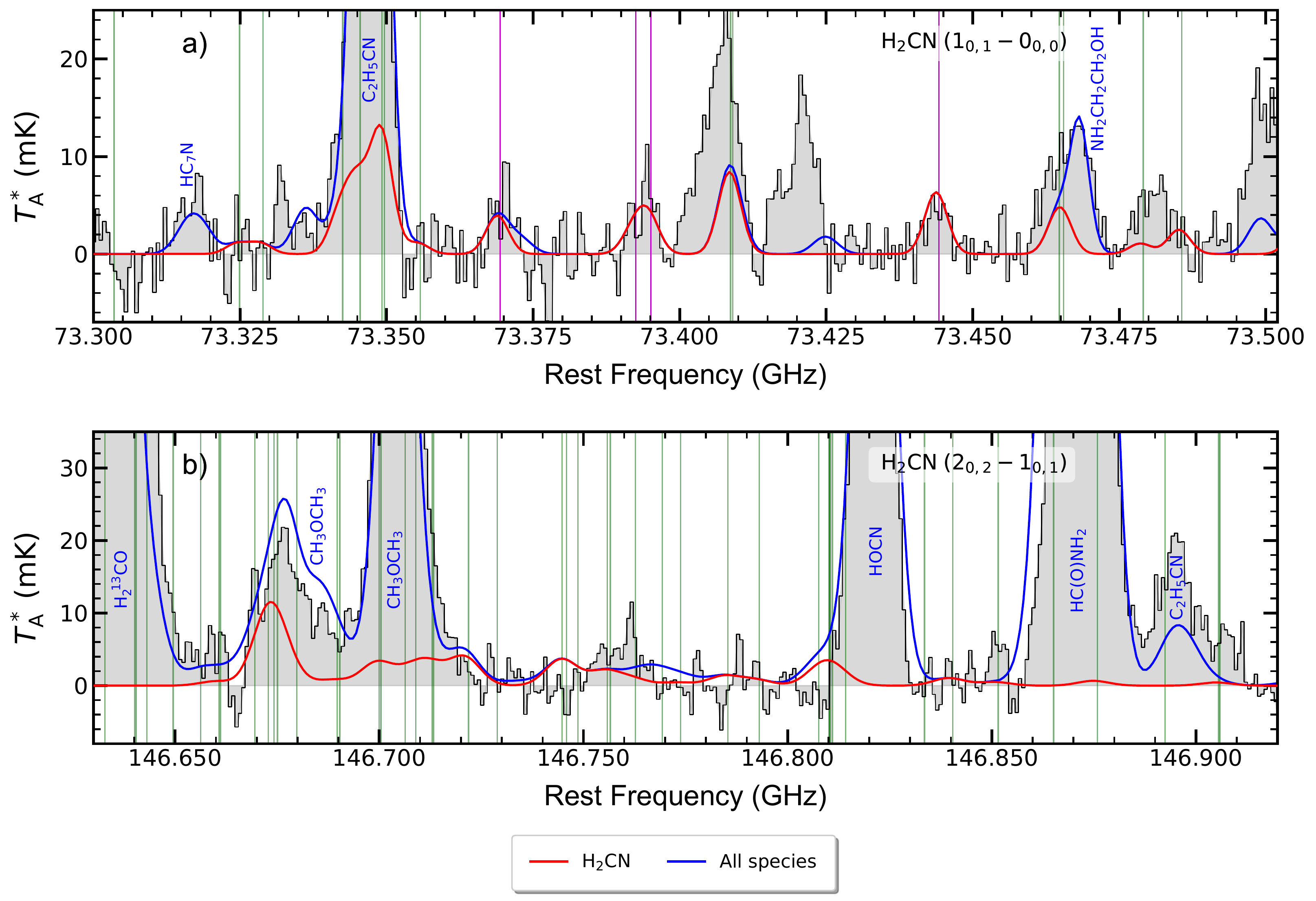}
    \caption{Same as Fig.~\ref{fig:H2NC_lines} but for \ce{H2CN}. Green vertical lines indicate other \ce{H2CN} transitions that were not used in the fit (see text). The spectroscopic information for all of the observed transitions is given in Table~\ref{tab:H2CN_spectroscopy} of Appendix~\ref{sec:spectroscopy_tables}.}
    \label{fig:H2CN_lines}
\end{figure*}

The hyperfine structure of the $1_{01} - 0_{00}$ and $2_{02} - 1_{01}$ rotational transitions has also been detected for \ce{H2CN} for the first time towards G+0.693, as shown in Fig.~\ref{fig:H2CN_lines}. The two \ce{H2CN} rotational transitions observed present upper energy levels of 3.5\K and 10.6\K, respectively (upper and bottom panels of Fig.~\ref{fig:H2CN_lines}). As in the case of \ce{H2NC}, only transitions with intensity $T_\text{A}^* > \sigma/3$ are considered to compute the synthetic spectrum. The spectroscopic information of \ce{H2CN} detected lines can be found in Table~\ref{tab:H2CN_spectroscopy} of Appendix~\ref{sec:spectroscopy_tables}. 

In contrast to the \ce{H2NC} detected lines (Fig.~\ref{fig:H2NC_lines}), most of \ce{H2CN} transitions are highly blended with transitions from other species already identified, mainly \ce{C2H5CN}, \ce{NH2CH2CH2OH} and \ce{CH3OCH3} (see Fig.~\ref{fig:H2CN_lines}). Four hyperfine \ce{H2CN} transitions at 73.369, 73.393, 73.395 and 73.444\GHz are completely unblended (magenta vertical lines in Fig.~\ref{fig:H2CN_lines}), and we have used them to perform AUTOFIT. Since all of them have the same energy levels, the AUTOFIT procedure does not converge when leaving free the $T_\text{ex}$. We have checked how different assumed values of $T_\text{ex}$ affect the LTE synthetic spectrum of the higher-energy $2_{02} - 1_{01}$ lines. We found that $T_\text{ex}$ should be indeed close to 3\K, similarly to \ce{H2NC}, since higher temperatures clearly overestimate the $2_{02} - 1_{01}$ transitions (see Fig.~\ref{fig:H2CN_tex} of Appendix~\ref{sec:H2CN_tex}). Hence, we have confidently used the $T_\text{ex}$ estimated for \ce{H2NC} (3.28\K) to perform the fit of \ce{H2CN}. Similarly, we also fixed $v_\text{LSR}$ to 71.0\kms and the $\text{FWHM}$ to 18.3\kms, as derived for \ce{H2NC}. The AUTOFIT procedure derives $N = (9.2\pm1.1) \times 10^{12} \, \text{cm}^{-2}$. The corresponding \ce{H2CN} molecular abundance relative to \ce{H2} is (6.8$\pm$1.3) $\times 10^{-11}$  (see Table~\ref{tab:fitting_parameters}).

\subsection{The \ce{H2CN}/\ce{H2NC} abundance ratio towards G+0.693}

\setlength{\tabcolsep}{2.6pt} 

\begin{table}
    \centering
    \caption{Physical parameters derived by the LTE best fit for \ce{H2CN} and \ce{H2NC}. The parameters assumed in the fit procedure are shown without errors, as explained in Sect.~\ref{sec:H2NC_detection} for \ce{H2NC} and Sect.~\ref{sec:H2CN_detection} for \ce{H2CN}. The fractional abundances with respect to \ce{H2} are shown in the last column. Uncertainties correspond to $1\sigma$ standard deviations, except for $N_\text{\ce{H2}}$ which is assumed to be 15$\%$ of its value.}
    \label{tab:fitting_parameters}
    \begin{tabular}{c c c c c c}
        \hline
        Molecule & $N$ & $T_\text{ex}$ & $\text{FWHM}$ & $v_\text{LSR}$ & $N/N_{\text{\ce{H2}}}$ \\
        \rule{0pt}{2.5ex}   
        & $(\times 10^{12} \, \text{cm}^{-2})$ & $(\text{K})$ & $(\text{km} \, \text{s}^{-1})$ & $(\text{km} \, \text{s}^{-1})$ & $(\times 10^{-11})$ \\
        \hline
        \ce{H2NC} & $4.2 \pm 0.7$ & $3.28 \pm 0.11$ & $18.3 \pm 1.1$ & $71.0 \pm 0.5$ & $3.1 \pm 0.7$\\
        \ce{H2CN} & $9.2 \pm 1.1$ & $3.28$ & $18.3$ & $71.0$ & $6.8 \pm 1.3$ \\
        \hline
        \multicolumn{6}{p{.45\textwidth}}{\footnotesize \textbf{Notes.} We have used the $N_{\text{\ce{H2}}}$ value derived by \citep{Martin2008} of 1.35$\times10^{23} \, \text{cm}^{-2}$.}
    \end{tabular}
\end{table}

Based on the column density estimates, we have obtained a \ce{H2CN}/\ce{H2NC} ratio of 2.2$\pm$0.5 towards G+0.693. This ratio is well above 1, consistent within uncertainty to what was found for the warm gas towards the galaxy at $z = 0.89$ in front of PKS 1830-211 quasar (3.7$\pm$1.1, \citealt{Cabezas2021}). In our analysis, the isomeric ratio has been derived using the same $T_\text{ex}$ for both molecules (3.28\K), as \citet{Cabezas2021} and \citet{Agundez2023} also did for the other sources. We note that this low excitation temperature is very similar to that derived by the former authors for L483 and B1-b. 
We have also evaluated how the \ce{H2CN}/\ce{H2NC} ratio varies assuming different $T_\text{ex}$ values for \ce{H2CN}, while keeping fixed the $T_\text{ex}$ of \ce{H2NC} to 3.28\K as derived from its fit. Since we have already demonstrated that the $T_\text{ex}$ of \ce{H2CN} should be close to 3\K and cannot be >4\K (as indicated in Fig.~\ref{fig:H2CN_tex} of Appendix~\ref{sec:H2CN_tex}), we have varied $T_\text{ex}$ within the $2\sigma$ confidence level of the fit of \ce{H2NC}, i.e., $\sim$3.0$-$3.5\K. We have found that the \ce{H2CN}/\ce{H2NC} abundance ratio ranges between 1.7 and 3.0, remaining unchanged with respect to the original estimate and errors.

\section{Discussion}\label{sec:discussion}

\subsection{The \ce{H2CN}/\ce{H2NC} abundance ratio behaviour with $T_\text{kin}$}\label{sec:H2CN/H2NC_vs_Tkin}

The G+0.693 molecular cloud is the first warm galactic molecular cloud and the second warm interstellar source, after the $z=0.89$ galaxy in front of PKS 1830-211 quasar (\citealt{Tercero2020}), where both the \ce{H2CN} and \ce{H2NC} isomers have been detected. The \ce{H2CN}/\ce{H2NC} ratio we have derived towards G+0.693, 2.2$\pm$0.5, along with that previously derived for PKS 1830-211 ($3.7 \pm 1.1$), confirms that this isomeric ratio in warm gas is higher than that found in cold dense clouds ($T_\text{kin} \sim 10$\K), where it is $\sim$1. 

\begin{figure}
    \centering
    \includegraphics[width=\columnwidth]{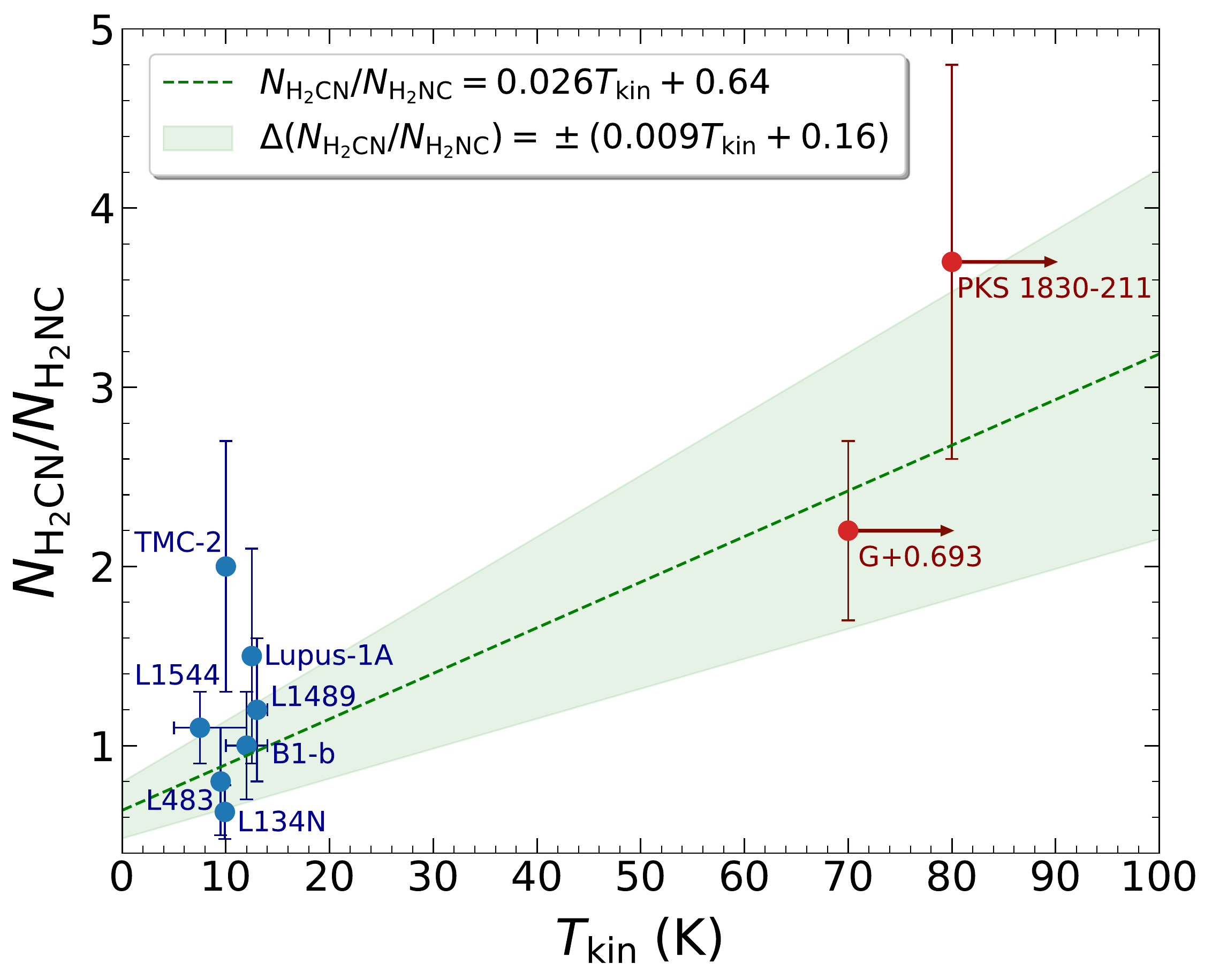}
    \caption{\ce{H2CN}/\ce{H2NC} abundance ratio versus $T_\text{kin}$. Blue and red points are associated to cold ($T_\text{kin}\sim10$\K) and warm ($T_\text{kin} > 70$\K) sources. The dashed green line traces the linear fit shown in eq.~(\ref{eq:ratio_correlation}), while the green shaded region encompasses the 1$\sigma$ of uncertainty on the linear regression fit. The $T_\text{kin}$ of G+0.693 is assumed to be $>$70\K from \ce{CH3CN} observations performed by \citet{Zeng2018}. The rest of $T_\text{kin}$ values were taken from \citet{Anglada1997} and \citet{Tafalla2000} for L483, from \citet{Marcelino2005} for B1-b, from \citet{Henkel2008} for PKS 1830-211 and from \citet{Benson&Myers1983}, \citet{Dickens2000}, \citet{Benedettini2012}, \citet{Wu2019} and \citet{Bianchi2023} for TMC-2, L134N, Lupus-1A, L1489 and L1544, respectively. The \ce{H2CN}/\ce{H2NC} ratios of the sources not studied in this work were taken from \citet{Cabezas2021} and \citet{Agundez2023}.}
    \label{fig:H2CN/H2NC_ratio}
\end{figure}

We show in Fig.~\ref{fig:H2CN/H2NC_ratio} the dependence of the \ce{H2CN}/\ce{H2NC} ratio with $T_\text{kin}$. For G+0.693, we have considered $T_\text{kin} > 70$\K, since \citet{Zeng2018} found that $T_\text{kin}$ ranges between $\sim$70\K and $\sim$140\K by analysing the $K$-ladders of different rotational transitions of \ce{CH3CN}, which is in good agreement with the values derived using \ce{NH3} towards Galactic Centre molecular clouds (\citealt{Guesten1985}; \citealt{Huettemeister1993}; \citealt{Krieger2017}). In the case of PKS 1830-211, we have adopted the lower limit of $\gtrsim$80\K derived by \citet{Henkel2008} using \ce{NH3} absorption.  

To compute the uncertainties of the isomeric ratios for which these were not provided, we have used the integrated intensity uncertainties resulting from the published line fitting. For each cold source, we have used the following $T_\text{kin}$: 9.5$\pm$0.3\K for L483 (from \ce{NH3} and \ce{CO} emission, \citealt{Anglada1997} and \citealt{Tafalla2000}, respectively), 12$\pm$2\K for B1-b (from \ce{CO} emission, \citealt{Bachiller&Cernicharo1984}; \citealt{Marcelino2005}), $\sim$10\K for TMC-2 (from \ce{HC5N} emission, \citealt{Benson&Myers1983}), 9.9$\pm$0.5\K for of L134N (from \ce{NH3} emission, \citealt{Dickens2000}), 12.5$\pm$0.2\K for Lupus-1A (based also on \ce{NH3} emission, \citealt{Benedettini2012}), 13$\pm$1\K for L1489 (using \ce{CH3CCH} emission, \citealt{Wu2019}), and  7.5\K with a 1$\sigma$ confidence range of 5$-$12\K for L1544 (as \citealt{Bianchi2023} estimated from \ce{HC5N}, \ce{HC7N} and \ce{HC9N} emission).

Fig.~\ref{fig:H2CN/H2NC_ratio} hints a clear dependence between the \ce{H2CN}/\ce{H2NC} abundance ratio and $T_\text{kin}$, with ratios above 2 for the warmer sources (G+0.693 and PKS 1830-211). On the other hand, the average value of this ratio found towards the seven cold clouds is 1.2$\pm$0.6. The ratios derived for these objects range between 0.6 and 1.5, with the only exception of TMC-2 (Fig.~\ref{fig:H2CN/H2NC_ratio}), which shows the largest uncertainty.

Therefore, the \ce{H2CN}/\ce{H2NC} ratio increases at higher $T_\text{kin}$, similar to what was found for the \ce{HCN}/\ce{HNC} abundance ratio (e.g., \citealt{Hacar2020}). Considering a $T_\text{kin}$ of 70\K for G+0.693, we performed a weighted linear regression fit, which is presented with a green dashed line in Fig.~\ref{fig:H2CN/H2NC_ratio} and accounts for uncertainties, yielding:

\begin{equation}
    \frac{\ce{H2CN}}{\ce{H2NC}} \equiv \frac{N_{\ce{H2CN}}}{N_{\ce{H2NC}}} = \left(0.026 \pm 0.009\right) T_\text{kin} + \left(0.64 \pm 0.16\right).
    \label{eq:ratio_correlation}
\end{equation}

In contrast to the $\ce{HCN}/\ce{HNC}$ ratio relation with $T_\text{kin}$ found by \citet{Hacar2020} for \ce{HCN}/\ce{HNC} ratios $\leq$4 ($T_\text{kin} = 10 \times \ce{HCN}/\ce{HNC}$ or, equivalently, $\ce{HCN}/\ce{HNC} = 0.1 \, T_\text{kin}$), the correlation we have found for the \ce{H2CN}/\ce{H2NC} ratio has a shallower slope of $0.026$. This indicates that the \ce{H2CN}/\ce{H2NC} ratio, although dependent on $T_\text{kin}$, seems to be less sensitive to it than the \ce{HCN}/\ce{HNC} ratio. As this value is in fact an upper limit for the slope, these conclusions still hold if a $T_\text{kin}$ of $\sim$140\K (the highest value in the temperature range derived by \citealt{Zeng2018}) is adopted for G+0.693, in which case a slope of 0.012 is obtained. 
In any case, we note that the trend found for the \ce{H2CN}/\ce{H2NC} ratio is still based on only two warm interstellar sources, and more detections of this isomeric pair towards other regions, covering sources with high ($T_\text{kin}>70$\K), and also intermediate temperatures (10\K $<T_\text{kin}<$ 70\K), would help to better describe the trend with the temperature.

\subsection{Interstellar chemistry of \ce{H2CN} and \ce{H2NC}}

The origin of the observed \ce{H2CN}/\ce{H2NC} abundance ratios and their possible link with the gas $T_\text{kin}$ can be related to different physical conditions or chemical processes, which can take place both in the gas phase and/or in the ices covering the interstellar dust grains. For example, it has been shown that the \ce{HCN}/\ce{HNC} ratio might depend on the collisional excitation of these molecules with \ce{He} (\citealt{Sarrasin2010}, \citealt{Hays2022}), on the increase of the cosmic-ray ionization rate in high temperature environments ($\sim$100\K, \citealt{Behrens2022}), on the isomerization of \ce{HNC} into \ce{HCN} in the icy dust grains surfaces \citep{Baiano2022}, on the ultraviolet radiation field \citep{Bublitz2022}, and on the kinetic temperature (\citealt{Jin2015}; \citealt{Colzi2018}; \citealt{Hacar2020}; \citealt{Pazukhin2022}). Motivated by the trend found in Sect.~\ref{sec:H2CN/H2NC_vs_Tkin}, we focus here on the possible dependence of the \ce{H2CN}/\ce{H2NC} ratio with $T_\text{kin}$, discussing possible chemical reactions forming and destroying these two radicals. 

\begin{figure}
    \centering
    \includegraphics[width=\columnwidth]{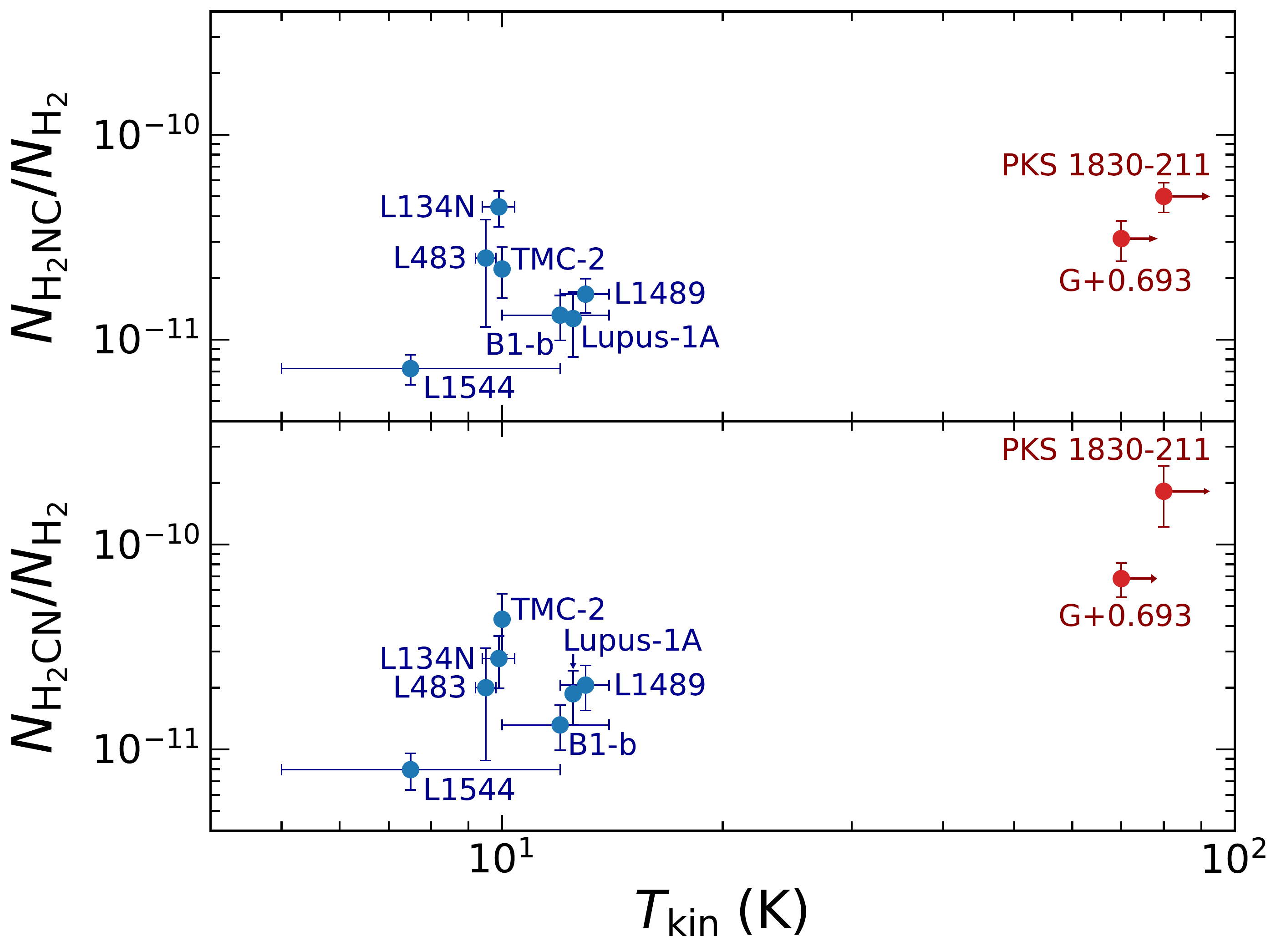}
    \caption{\ce{H2NC} (upper panel) and \ce{H2CN} (lower panel) molecular abundances relative to \ce{H2} versus the gas kinetic temperature. Blue and red colors are associated to the cold ($T_\text{kin}\sim10$\K) and warm ($T_\text{kin} > 70$\K) sources, respectively. 
    \ce{H2CN} and \ce{H2NC} molecular abundances come from this work and \citet{Cabezas2021} and \citet{Agundez2023}.
    The \ce{H2} column density of G+0.693 is $(1.4\pm0.2)\times10^{23} \, \text{cm}^{-2}$ \citep{Martin2008}, while for the rest of the sources are: $(2.2\pm0.3)\times10^{22} \, \text{cm}^{-2}$ (PKS 1830-211; \citealt{Muller2014}), $(4.0\pm2.0)\times10^{22} \, \text{cm}^{-2}$ (L483; \citealt{Agundez2019}), $(7.6\pm1.1)\times10^{22} \, \text{cm}^{-2}$ (B1-b; \citealt{Daniel2013}), $(9.5\pm1.4)\times10^{21} \, \text{cm}^{-2}$ (TMC-2; \citealt{Turner1997}), $(9.0\pm1.4)\times10^{21} \, \text{cm}^{-2}$ (L134N; \citealt{Dickens2000}), $(1.50\pm0.23)\times10^{22} \, \text{cm}^{-2}$ (Lupus-1A; \citealt{Agundez2015}), $(1.02\pm0.07)\times10^{22} \, \text{cm}^{-2}$ (L1489; \citealt{Wu2019}) and $(5.4\pm0.8)\times10^{22} \, \text{cm}^{-2}$ (L1544; \citealt{Jimenez-Serra2016}). When not given, uncertainties of $N$(H$_2$) were adopted to be 15\% of the value.}
    \label{fig:H2CN/H2NC_H2_abundances}
\end{figure}

Unlike \ce{H2CN}, the chemistry of \ce{H2NC} is very poorly explored, since it is missing in chemical kinetic databases commonly used in astrochemistry, such as UDfA (UMIST Database for Astrochemistry; \citealt{UDfA}) or KIDA (KInetic Database for Astrochemistry; \citealt{KIDA}). Moreover, the predominantly detection of these two isomers in cold clouds, which are expected to be mainly affected by gas-phase chemistry, have only triggered the study of different gas-phase chemical routes. Nonetheless, chemical pathways occurring on the ices covering dust grains can be especially important in the case of G+0.693, where the presence of shocks is able to inject the molecules into the gas phase; or in warm sources such as PKS 1830-211, where thermal and shock-induced desorption are also possible (\citealt{Tercero2020}; \citealt{Muller2021}). To explore this hypothesis, we have computed the molecular abundances of both \ce{H2CN} and \ce{H2NC} isomers with respect to \ce{H2} for all of the sources, and analysed their dependence on the gas $T_\text{kin}$ (see Fig.~\ref{fig:H2CN/H2NC_H2_abundances}). The \ce{H2} column densities of those objects not studied in this work where obtained from various studies (\citealt{Muller2014} for PKS 1830-211, \citealt{Agundez2019} for L483, \citealt{Daniel2013} for B1-b, \citealt{Turner1997} for TMC-2, \citealt{Dickens2000} for L134N, \citealt{Agundez2015} for Lupus-1A, \citealt{Wu2019} for L1489 and \citealt{Jimenez-Serra2016} for L1544), and range from $\sim$$(1-8)\times10^{22} \, \text{cm}^{-2}$. An uncertainty of 15\% of their value was assumed when not provided, as done for G+0.693. In the case of TMC-2 and L134N, \ce{H2} volume density was converted into column density by assuming spherical geometry and using the source size provided by the corresponding above-mentioned works. 

As Fig.~\ref{fig:H2CN/H2NC_H2_abundances} illustrates, both the warm/shocked sources (G+0.693 and PKS 1830-211) exhibit  higher abundances of \ce{H2CN}, which are up to an order of magnitude larger than those observed in most of the cold clouds. \ce{H2NC} appears also to be more abundant in these two warm objects, albeit the difference with respect to the cold sources is not as significant. 
These findings suggest that surface chemistry might be enhancing both \ce{H2NC} and, more notably, \ce{H2CN} abundances in warm/shocked environments.

In the next subsections we discuss the possible gas-phase formation and destruction routes that have been studied so far for \ce{H2CN}
(\citealt{Cimas&Largo2006}; \citealt{Holzmeier2013}; \citealt{Bourgalais2015}), and \ce{H2NC} (\citealt{Cabezas2021,Agundez2023}). Moreover, we also propose alternative pathways that might occur on the surface of the ices, which can potentially enhance their abundances in warm and/or shocked regions.

\subsubsection{\ce{H2CN} and \ce{H2NC} formation mechanisms}

A first \ce{H2CN} formation route occurs through the gas-phase reaction $\ce{CH3 + N(^4S)} \rightarrow \ce{H2CN + H}$, which was studied experimentally by \citet{Marston1989}, and also computationally by \citet{Cimas&Largo2006} through calculations at the CCSD(T)/cc-pVTZ, MP2/CBS, B3LYP/CBS and G2 levels of theory. The latter authors showed that this reaction first gives \ce{H3CN} as an intermediate species, which subsequently forms \ce{H2CN} with a branching ratio of $\sim$0.997\footnote{Modification of the temperature in the range 100$-$500\K alters these branching ratios only in the fourth decimal figure towards the formation of \ce{H2CN} and \ce{trans-HCNH}, but remains constant for \ce{H2NC}.}. They also considered the formation of \ce{H2NC} (with a negligible branching ratio of $\sim$10$^{-5}$) and \textit{trans}-\ce{HCNH} (branching ratio of $\sim$0.0025). All these processes are exothermic and barrierless, and \ce{H2CN} is the main product being the most favourable pathway both thermodynamically and kinetically. \citet{Cimas&Largo2006} also analysed the formation of \ce{HCN} (branching ratio of $\sim$0.185\footnote{Modification of the temperature in the range 100$-$500\K alters these branching ratios only in the third decimal figure.}) and \ce{HNC} ($\sim$0.071) through the same reaction, but starting with atomic \ce{N} in an excited state \ce{(^2D)}. Although the formation of \ce{HCN} and \ce{HNC} is thermodinamically favoured, these molecules are not the preferred products from the kinetic point of view, as their transition states lie higher in energy than those for \ce{H2CN}. Therefore, \ce{H2CN} is the main product of this reaction with a branching ratio of $\sim$0.745, although the formation of \ce{HCN} and \ce{HNC} is also possible. However, this reaction will not occur in the ISM since the population of excited \ce{N(^2D)} atoms is going to be negligible with respect to that of \ce{N(^4S)} atoms.

Another possible \ce{H2CN} formation route involves the gas-phase reaction of $\ce{C + NH3} \rightarrow \ce{H2CN + H}$, which was initially studied by \citet{Bourgalais2015} through both experiments and theoretical calculations at the MRCI+Q/aug-cc-pVQZ, CCSD(T)/aug-cc-pVQZ and M06-2X/cc-pVQZ levels. These authors reported that the \ce{H2CN + H} exit channel represents about $100\%$ of the product yield for this reaction (branching ratio $\simeq$1), being an exothermic reaction that occurs without an emerged barrier. They also showed that the exothermic channels that lead to \ce{HCN + 2 H} or \ce{HNC + 2 H} are final minor products. However, although these authors did not discuss the formation of \ce{H2NC}, \citet{Cabezas2021} argued that it could also take place through the \ce{C + NH3} reaction. This is because it first involves the formation of the \ce{CNH3} intermediate species in a barrierless process, which followed by the elimination of a \ce{H} atom would directly lead to \ce{H2NC + H} (similarly to the most favoured path of the \ce{CH3 + N} reaction from \citealt{Cimas&Largo2006}, which first involves the formation of \ce{H3CN}). Moreover, the chemical structure of the intermediate species \ce{CNH3} could also favour the formation of \ce{H2NC} instead of \ce{H2CN}, in which significant atomic rearrangement is required. More recently, \citet{Agundez2023} revisited this reaction theoretically and found that both \ce{H2CN} and \ce{H2NC} can be formed in an exothermic process. Their calculations using the RCCSD(T)-F12, MRCI-F12 and MRCI-F12+Q methods in combination with the cc-pVQZ-F12 and cc-pVTZ-F12 basis sets, showed that the \ce{CNH3} intermediate species could first evolve towards two transition states very close in energy, but located below the reactants. The highest energy transition state would directly lead to \ce{H2NC + H} products in a barrierless process, while the \ce{H2CN + H} exit channel would be favoured by means of a statistical mechanism starting from the lowest energy transition state, and consequently going through different reaction intermediates. 
These authors predicted that the \ce{H2CN}/\ce{H2NC} ratio at low temperatures (10\K) has a value between 1.25 and 2, reasonably similar to the values observed in the cold molecular clouds (see Fig. \ref{fig:H2CN/H2NC_ratio}). Notwithstanding, the absence of an entrance barrier points to this reaction not being temperature dependent, making it unable to explain the higher ratios and molecular abundances found towards the warm sources. Although it might be a dominant formation route of both isomers in the gas phase, it is not responsible for the \ce{H2CN}/\ce{H2NC} abundance ratio dependence on temperature found.

\citet{Talbi1996} also studied the formation of \ce{H2CN} and \ce{H2NC} via \ce{HCN + H} and \ce{HNC + H} gas-phase reactions. However, in addition to the fact that forming a single product from a two-body reaction makes it highly unstable, their high activation barriers of $\sim$15.2\kcalmol ($\sim$7700\K) and $\sim$19.2\kcalmol ($\sim$9700\K) make these processes highly inefficient at the conditions of the ISM. 

The detection of \ce{H2CN} and \ce{H2NC} towards the G+0.693 molecular cloud, whose chemistry is dominated by shock-induced desorption (\citealt{Rivilla2020,Rivilla2021a, Rivilla2022c}), led us to consider also grain surface chemistry. Possible formation routes might be one H-atom addition through the direct hydrogenation of \ce{HCN} and \ce{HNC}, respectively, on the surface of dust grains. \citet{Theule2011} performed hydrogenation experiments of pure \ce{HCN} ices, and found that they fully hydrogenates into methylamine (\ce{CH3NH2}). In this process, \ce{H2CN} is a non-stable intermediate species. However, we note that the H-atom flux in the experiments is likely much higher than that in the ISM conditions, where the survival of intermediate species is then favoured, although at much lower abundances than the parent species. In G+0.693, the \ce{HCN}/\ce{H2CN} ratio is indeed (5.0$\pm$0.6)$\times10^3$ (using the column density reported here and in \citealt{Colzi2022}), in agreement with this hypothesis. In a similar way, \ce{H2NC} might be formed on grains through the hydrogenation of \ce{HNC}, but to our best knowledge there are no laboratory experiments that have specifically studied this proccess. The detections of \ce{H2CN} and \ce{H2NC} in shock-dominated and/or thermal-dominated sources like G+0.693 and PKS 1830-211 impose the need of exploring in more detail its chemistry on the surface of ices, both in the laboratory and also by dedicated quantum chemical calculations.

\subsubsection{\ce{H2CN} and \ce{H2NC} destruction mechanisms}

Given that it has been proposed that the dependence of the \ce{HCN}/\ce{HNC} ratio with $T_\text{kin}$ is mainly due to destruction and isomerization mechanisms (\citealt{Lin1992}; \citealt{Talbi1996}; \citealt{Graninger2014}; \citealt{Jin2015}; \citealt{Hacar2020}), it is also worth to explore possible destruction pathways for the \ce{H2CN} and \ce{H2NC} isomers. However, the destruction of these two isomers has been poorly studied so far. In the case of \ce{H2CN}, to our best knowledge, the only destruction routes proposed in the literature are gas-phase reactions with neutral \ce{H}, \ce{C}, \ce{N}, and \ce{O} atoms, which were summarised by \citet{Loison2014}, based on previous works by \citet{Fikri2001}, \citet{Cho&Andrews2011} or \citet{Hebrard2012} among others.

The \ce{H2CN + H} destruction reaction was first studied experimentally by \citet{Nesbitt1990}, who derived a lower limit for its rate constant. Through \textit{ab-initio} calculations at the M06-2X/cc-pVTZ and MRCI+Q/cc-pVQZ levels, \citet{Hebrard2012} determined that this reaction can give \ce{HCN + H2} or \ce{HNC + H2} as final products in an exothermic process without entrance barrier. Based on statistical calculations, these authors estimated that 20$\%$ of the total production goes into \ce{HNC} formation, while 80$\%$ into \ce{HCN}. 

Regarding the \ce{H2CN + N} reaction, \citet{Nesbitt1990} determined its rate constant (at temperatures in the range 200$-$363\K) with large uncertainties, so that extrapolation to lower temperatures would be unreliable. To form the products \ce{NH + HCN}, an emerged barrier of $\sim$1320\K should be overcome \citep{Cimas&Largo2006}, and it is not expected to occur under interstellar conditions. The preferable products of the reaction are then \ce{CH2 + N2} due to the absence of a barrier, as found by \citet{Hebrard2012} from \textit{ab-initio} calculations at the M06-2X/cc-pVTZ level.

The \ce{H2CN + C} and \ce{H2CN + O} reactions have not been specifically studied experimentally or theoretically. \citet{Loison2014} estimated their parameters by analysing similar chemical reactions (\citealt{Lau1999}; \citealt{Fikri2001}; \citealt{Zhang2001} ; \citealt{Koput2003}; \citealt{Osamura&Petrie2004}; \citealt{Cho&Andrews2011}). We note that special attention should be paid to \ce{H2CN + O} and \ce{H2NC + O} reactions, as it has been shown that the reaction \ce{HNC + O} could be the dominant mechanism producing the observed \ce{HCN}/\ce{HNC}, because the same reaction is unable to occur for \ce{HCN} under ISM temperature conditions \citep{Lin1992}. In a very similar way, the destruction reaction \ce{HNC + H} has been proposed to be an efficient mechanism converting \ce{HNC} into \ce{HCN} (\citealt{PineaudesForests1990}; \citealt{Talbi1996}; \citealt{Jin2015}; \citealt{Hacar2020}), so that the destruction reactions \ce{H2CN + H} and \ce{H2NC + H} might also deserve special consideration. Thus, taking the \ce{HNC} and \ce{HCN} isomeric pair as a reference, it would be interesting to analyse theoretically and experimentally the differences in \ce{H2CN} and \ce{H2NC} destruction reactions with atomic \ce{O} and \ce{H}, and their dependence with $T_\text{kin}$, which might be able to explain the observed trend of the \ce{H2CN}/\ce{H2NC} ratio (Sect.~\ref{sec:H2CN/H2NC_vs_Tkin}). 

Moreover, a temperature dependence of the gas-phase isomerization of \ce{H2NC} into \ce{H2CN} would be another way to explain the observed \ce{H2CN}/\ce{H2NC} abundance ratio dependence with $T_\text{kin}$. However, \citet{Holzmeier2013} argued that the isomerization cannot occur in either direction in gas-phase, due to the large difference in energy between these two isomers in their ground state ($\sim$12000\K or $\sim$24\kcalmol), and the even higher energy barriers within the isomerization process itself ($\sim$23000\K). These large barriers cannot be overcome under typical molecular ISM conditions ($T_\text{kin}$ $\sim$10$-$100\K), even considering quantum tunneling effects. Notwithstanding, tunneling probability strongly depends on the shape and width of the reaction barriers, so that a more detailed analysis is required. In any case, a very high degree of atom rearrangement is needed to form \ce{H2NC} from \ce{H2CN} and vice versa, which also makes isomerization very unlikely.

Regarding surface chemistry, as noted above, reactions involving \ce{H2NC} from \ce{H2CN} are poorly studied. Besides the H-atom addition mentioned before, which would destroy them to form more saturated species such as methanimine and methylamine (as shown in the experiments by \citealt{Theule2011}), it would be worth to study the possible surface isomerisation $\ce{H2CN} \rightleftharpoons \ce{H2NC}$. \citet{Baiano2022} demonstrated that a similar surface isomerisation process between \ce{HNC} and \ce{HCN} is possible in the presence of water molecules. They formed \ce{HCN} from \ce{HNC} in the presence of \ce{H2O} molecules through quantum tunneling effects, with a reaction rate that increases with the temperature. If an analogous process is also possible for \ce{H2CN} and \ce{H2NC}, this would already increase the \ce{H2CN}/\ce{H2NC} isomeric ratio on ices, which would be inherited by the gas phase after desorption (both shock-induced or thermal). This mechanism, which needs to be studied, might be able to naturally explain the higher \ce{H2CN}/\ce{H2NC} ratio found towards G+0.693 and PKS 1830-211.

\section{Conclusions}\label{sec:conclusions}

We have reported for the first time the detection of \ce{H2CN} and its high-energy metastable isomer \ce{H2NC} towards a warm ($T_\text{kin} > 70$\K) galactic molecular cloud, G+0.693-0.027, which is located in the Galactic Centre of our Galaxy. Using IRAM 30m telescope observations, we have detected several hyperfine components of the $N_{K_\text{a}K_\text{c}} = 1_{01} - 0_{00}$ and $2_{02} - 1_{01}$ transitions.
We performed a LTE fit to the observed spectra, deriving $N = (4.2 \pm 0.7) \times 10^{12} \, \text{cm}^{-2}$ and $T_\text{ex} = 3.28 \pm 0.11$\K for \ce{H2NC}, and $N = (9.2\pm1.1) \times 10^{12} \, \text{cm}^{-2}$ when assuming $T_\text{ex} = 3.28$\K for \ce{H2CN}. Thereby, we have derived molecular abundances with respect to \ce{H2} of (6.8$\pm$1.3) $\times 10^{-11}$ for \ce{H2CN} and of (3.1$\pm$0.7) $\times 10^{-11}$ for \ce{H2NC}, and an isomeric abundance ratio of $\ce{H2CN}/\ce{H2NC} = 2.2\pm0.5$. 
This result, along with the ratio previously found  towards the $T_\text{kin} \gtrsim 80$\K gas of a $z = 0.89$ galaxy in front of the quasar PKS 1830-211 (3.7$\pm$1.1), confirms that the $\ce{H2CN}/\ce{H2NC}$ ratio is higher in warm sources than in cold  ($T_\text{kin} \sim 10$\K) clouds, where it is $\sim$1. This suggest that, similarly to the HCN/HNC ratio, the $\ce{H2CN}/\ce{H2NC}$ abundance ratio can also be used as a new potential temperature tracer for the ISM. We have found that the dependence of the \ce{H2CN}/\ce{H2NC} isomeric ratio with $T_\text{kin}$ is shallower ($\sim$0.026$T_\text{kin}$) than that previously found for the \ce{HCN}/\ce{HNC} ratio ($\sim$0.1$T_\text{kin}$).
The observed dependence of the \ce{H2CN}/\ce{H2NC} ratio with the gas $T_\text{kin}$ cannot be fully explained only with the current proposed formation and destruction pathways in the gas phase, although the latter should be studied in more detail. In particular, the higher ratios and molecular abundances (up to one order of magnitude larger in the case of \ce{H2CN}) found in G+0.693 (where shock-induced desorption dominates) and PKS 1830-211 (where thermal desorption is also possible), impose the need of considering also surface chemistry on the icy grain mantles.
Hence, the observational result presented here highlights the importance of performing new laboratory experiments and quantum chemical calculations to study, e.g., the surface formation of \ce{H2CN} and \ce{H2NC} through the direct hydrogenation of \ce{HCN} and \ce{HNC}, respectively, their destruction by subsequent H-atoms additions, or their possible isomerization on the ices, as previously proposed for \ce{HCN} and \ce{HNC}. These mechanisms can introduce an isomeric preference on ices, which later on would be inherited in the gas phase in those sources where shocks or heating can desorb the molecules. This might contribute to explain the higher  \ce{H2CN}/\ce{H2NC} values derived towards G+0.693 and PKS 1830-211, compared to cold molecular clouds where pure gas-phase chemistry is expected to dominate.

\section*{Acknowledgements}

We are very grateful to the IRAM 30m telescope staff for their precious help during the different observing runs. We acknowledge Germ\'an Molpeceres for enlightening discussions about the surface chemistry of the species studies in this work. D.S.A. acknowledges the funds provided by the Consejo Superior de Investigaciones Científicas (CSIC) and the Centro de Astrobiología (CAB) through the project 20225AT015 (Proyectos intramurales especiales del CSIC). L.C., I.J.-S., and J.M.-P. acknowledge financial support through the Spanish grant PID2019-105552RB-C41 funded by MCIN/AEI/10.13039/501100011033. V.M.R. has received support from the project RYC2020-029387-I funded by MCIN/AEI /10.13039/501100011033, and from the Comunidad de Madrid through the Atracci\'on de Talento Investigador Modalidad 1 (Doctores con experiencia) Grant (COOL: Cosmic Origins Of Life; 2019-T1/TIC-5379).

\section*{Data Availability}

The data underlying this article will be shared on reasonable request to vrivilla$@$cab.inta-csic.es .



\bibliographystyle{mnras}
\bibliography{biblio} 




\appendix

\section{Spectroscopic parameters of \ce{H2CN} and \ce{H2NC} excited transitions}
\label{sec:spectroscopy_tables}

Tables~\ref{tab:H2NC_spectroscopy} and \ref{tab:H2CN_spectroscopy} list the \ce{H2NC} and \ce{H2CN} identified transitions towards G+0.693 cloud, respectively. We show only transitions with peak intensities of $T_\text{A}^* > 0.41 \, \text{mK}$, which corresponds to $\sigma/3$ (where $\sigma$ is the rms of the spectra), according to the LTE fit performed with MADCUBA. The last column (named `Checked'), indicates if each particular transition has been used for the LTE fit (Yes) or not (No). All of the transitions listed in Tables~\ref{tab:H2NC_spectroscopy} and \ref{tab:H2CN_spectroscopy} also appear in Fig.~\ref{fig:H2NC_lines} (\ce{H2NC}) and Fig.~\ref{fig:H2CN_lines} (\ce{H2CN}), respectively.
Both tables provide the most relevant spectroscopic parameters for each transition, which are: the rest frequency $\nu_\text{obs}$, the quantum numbers of the levels involved in the transitions ($N$, $K_\text{a}$, $K_\text{c}$, $J$, $F_\text{1}$, and $F$), the energy of the upper level $E_\text{up}$, and their integrated intensity at 300\K ($I$). All of these transitions belong to ortho-\ce{H2NC} (Table~\ref{tab:H2NC_spectroscopy}) and ortho-\ce{H2CN} (Table~\ref{tab:H2CN_spectroscopy}), respectively. Spectroscopic information has been obtained from CDMS and \citet{Cabezas2021} for \ce{H2NC}, and from CDMS and \citet{Yamamoto1992} for \ce{H2CN}.

\setlength{\tabcolsep}{8pt} 

\begin{table*}
    \centering
    \caption{\label{tab:H2NC_spectroscopy} Main parameters of \ce{H2NC} transitions observed towards G+0.693 molecular cloud. The first column shows the rest frequency for each transition in units of $\text{GHz}$, while the next 12 gather the quantum numbers of the upper $\left(N', K_\text{a}', K_\text{c}', J', F_1', F'\right)$ and lower $\left(N'', K_\text{a}'', K_\text{c}'', J'', F_1'', F''\right)$ levels, obtained from  \citet{Cabezas2021} and CDMS. $\log I$ is the base 10 logarithm of the integrated intensity in units of $\text{nm}^2 \, \text{MHz}$ at a fixed temperature of 300\K, and $E_\text{up}$ is the energy in \K of the upper level involved in the transition. The last column (named 'Checked') indicates if each particular transition has been used to perform the AUTOFIT LTE fit (Yes) or not (No). All of the transitions belong to ortho-\ce{H2NC}.}
    \begin{tabular}{c c c c c c c c c c c c c c c c}
        \hline
        $\nu_\text{rest}$ & $N'$ & $K_\text{a}'$ & $K_\text{c}'$ & $J'$ & $F_1'$ & $F'$ & $N''$ & $K_\text{a}''$ & $K_\text{c}''$ & $J''$ & $F_1''$ & $F''$ & $\log I$ & $E_\text{up}$ & Checked \\
        ($\text{GHz}$) &  &  &  &  &  &  &  &  &  &  &  &  & ($\text{nm}^2 \, \text{MHz}$) & ($\text{K}$)  &  \\
        \hline
        72.188328 & 1 & 0 & 1 & $3/2$ & $3/2$ & $5/2$ & 0 & 0 & 0 & $1/2$ & $1/2$ & $3/2$ & -4.497 & 3.5 & Yes \\
        72.189077 & 1 & 0 & 1 & $3/2$ & $1/2$ & $3/2$ & 0 & 0 & 0 & $1/2$ & $1/2$ & $1/2$ & -4.832 & 3.5 & Yes \\
        72.191838 & 1 & 0 & 1 & $3/2$ & $5/2$ & $3/2$ & 0 & 0 & 0 & $1/2$ & $3/2$ & $3/2$ & -5.068 & 3.5 & Yes \\
        72.194219 & 1 & 0 & 1 & $3/2$ & $5/2$ & $7/2$ & 0 & 0 & 0 & $1/2$ & $3/2$ & $5/2$ & -4.071 & 3.5 & Yes \\
        72.197262 & 1 & 0 & 1 & $1/2$ & $1/2$ & $3/2$ & 0 & 0 & 0 & $1/2$ & $1/2$ & $1/2$ & -4.974 & 3.5 & Yes \\
        72.198211 & 1 & 0 & 1 & $3/2$ & $3/2$ & $1/2$ & 0 & 0 & 0 & $1/2$ & $3/2$ & $3/2$ & -5.064 & 3.5 & Yes \\
        72.199831 & 1 & 0 & 1 & $3/2$ & $1/2$ & $3/2$ & 0 & 0 & 0 & $1/2$ & $1/2$ & $3/2$ & -4.872 & 3.5 & Yes \\
        72.202571 & 1 & 0 & 1 & $3/2$ & $5/2$ & $3/2$ & 0 & 0 & 0 & $1/2$ & $3/2$ & $1/2$ & -4.471 & 3.5 & Yes \\
        72.206545 & 1 & 0 & 1 & $3/2$ & $1/2$ & $1/2$ & 0 & 0 & 0 & $1/2$ & $3/2$ & $3/2$ & -4.901 & 3.5 & Yes \\
        72.208502 & 1 & 0 & 1 & $3/2$ & $3/2$ & $5/2$ & 0 & 0 & 0 & $1/2$ & $3/2$ & $5/2$ & -4.498 & 3.5 & Yes \\
        72.208937 & 1 & 0 & 1 & $3/2$ & $3/2$ & $1/2$ & 0 & 0 & 0 & $1/2$ & $3/2$ & $1/2$ & -4.902 & 3.5 & Yes \\
        72.209450 & 1 & 0 & 1 & $3/2$ & $3/2$ & $3/2$ & 0 & 0 & 0 & $1/2$ & $3/2$ & $3/2$ & -4.472 & 3.5 & Yes \\
        72.210947 & 1 & 0 & 1 & $3/2$ & $5/2$ & $5/2$ & 0 & 0 & 0 & $1/2$ & $3/2$ & $3/2$ & -4.199 & 3.5 & Yes \\
        72.212383 & 1 & 0 & 1 & $1/2$ & $3/2$ & $5/2$ & 0 & 0 & 0 & $1/2$ & $1/2$ & $3/2$ & -4.498 & 3.5 & Yes \\
        72.217255 & 1 & 0 & 1 & $3/2$ & $1/2$ & $1/2$ & 0 & 0 & 0 & $1/2$ & $3/2$ & $1/2$ & -5.066 & 3.5 & Yes \\
        72.217855 & 1 & 0 & 1 & $1/2$ & $3/2$ & $3/2$ & 0 & 0 & 0 & $1/2$ & $1/2$ & $1/2$ & -4.768 & 3.5 & Yes \\
        72.219996 & 1 & 0 & 1 & $3/2$ & $1/2$ & $3/2$ & 0 & 0 & 0 & $1/2$ & $3/2$ & $5/2$ & -4.846 & 3.5 & Yes \\
        72.220178 & 1 & 0 & 1 & $3/2$ & $3/2$ & $3/2$ & 0 & 0 & 0 & $1/2$ & $3/2$ & $1/2$ & -5.072 & 3.5 & Yes \\
        72.222854 & 1 & 0 & 1 & $1/2$ & $1/2$ & $1/2$ & 0 & 0 & 0 & $1/2$ & $1/2$ & $3/2$ & -4.705 & 3.5 & Yes \\
        72.225310 & 1 & 0 & 1 & $1/2$ & $3/2$ & $1/2$ & 0 & 0 & 0 & $1/2$ & $1/2$ & $1/2$ & -4.704 & 3.5 & Yes \\
        72.228155 & 1 & 0 & 1 & $1/2$ & $1/2$ & $3/2$ & 0 & 0 & 0 & $1/2$ & $3/2$ & $5/2$ & -4.557 & 3.5 & Yes \\
        72.228583 & 1 & 0 & 1 & $1/2$ & $3/2$ & $3/2$ & 0 & 0 & 0 & $1/2$ & $1/2$ & $3/2$ & -4.605 & 3.5 & Yes \\
        72.232548 & 1 & 0 & 1 & $1/2$ & $3/2$ & $5/2$ & 0 & 0 & 0 & $1/2$ & $3/2$ & $5/2$ & -4.501 & 3.5 & Yes \\
        144.361610 & 2 & 0 & 2 & $5/2$ & $5/2$ & $7/2$ & 1 & 0 & 1 & $3/2$ & $3/2$ & $5/2$ & -3.486 & 10.4 & Yes \\
        144.362101 & 2 & 0 & 2 & $5/2$ & $3/2$ & $5/2$ & 1 & 0 & 1 & $3/2$ & $1/2$ & $3/2$ & -3.698 & 10.4 & Yes \\
        144.362392 & 2 & 0 & 2 & $5/2$ & $7/2$ & $9/2$ & 1 & 0 & 1 & $3/2$ & $5/2$ & $7/2$ & -3.301 & 10.4 & Yes \\
        144.366915 & 2 & 0 & 2 & $5/2$ & $3/2$ & $1/2$ & 1 & 0 & 1 & $3/2$ & $1/2$ & $1/2$ & -4.305 & 10.4 & Yes \\
        144.369440 & 2 & 0 & 2 & $5/2$ & $7/2$ & $5/2$ & 1 & 0 & 1 & $3/2$ & $5/2$ & $3/2$ & -3.558 & 10.4 & Yes \\
        144.369970 & 2 & 0 & 2 & $5/2$ & $5/2$ & $3/2$ & 1 & 0 & 1 & $3/2$ & $3/2$ & $1/2$ & -3.919 & 10.4 & Yes \\
        144.373607 & 2 & 0 & 2 & $5/2$ & $3/2$ & $5/2$ & 1 & 0 & 1 & $3/2$ & $3/2$ & $5/2$ & -4.183 & 10.4 & Yes \\
        144.375925 & 2 & 0 & 2 & $5/2$ & $5/2$ & $7/2$ & 1 & 0 & 1 & $3/2$ & $5/2$ & $7/2$ & -4.163 & 10.4 & Yes \\
        144.376393 & 2 & 0 & 2 & $5/2$ & $5/2$ & $3/2$ & 1 & 0 & 1 & $3/2$ & $5/2$ & $3/2$ & -4.200 & 10.4 & Yes \\
        144.376393 & 2 & 0 & 2 & $3/2$ & $3/2$ & $5/2$ & 1 & 0 & 1 & $1/2$ & $3/2$ & $5/2$ & -4.326 & 10.4 & Yes \\
        144.376968 & 2 & 0 & 2 & $5/2$ & $3/2$ & $3/2$ & 1 & 0 & 1 & $3/2$ & $3/2$ & $3/2$ & -4.183 & 10.4 & Yes \\
        144.377018 & 2 & 0 & 2 & $5/2$ & $5/2$ & $5/2$ & 1 & 0 & 1 & $3/2$ & $5/2$ & $5/2$ & -4.173 & 10.4 & Yes \\
        144.377058 & 2 & 0 & 2 & $5/2$ & $7/2$ & $7/2$ & 1 & 0 & 1 & $3/2$ & $5/2$ & $5/2$ & -3.401 & 10.4 & Yes \\
        144.377058 & 2 & 0 & 2 & $3/2$ & $5/2$ & $7/2$ & 1 & 0 & 1 & $1/2$ & $3/2$ & $5/2$ & -3.490 & 10.4 & Yes \\
        144.378508 & 2 & 0 & 2 & $5/2$ & $5/2$ & $5/2$ & 1 & 0 & 1 & $3/2$ & $3/2$ & $3/2$ & -3.645 & 10.4 & Yes \\
        144.379862 & 2 & 0 & 2 & $5/2$ & $3/2$ & $3/2$ & 1 & 0 & 1 & $3/2$ & $1/2$ & $1/2$ & -3.934 & 10.4 & Yes \\
        144.380762 & 2 & 0 & 2 & $3/2$ & $3/2$ & $5/2$ & 1 & 0 & 1 & $1/2$ & $1/2$ & $3/2$ & -3.907 & 10.4 & Yes \\
        144.383214 & 2 & 0 & 2 & $3/2$ & $3/2$ & $3/2$ & 1 & 0 & 1 & $1/2$ & $1/2$ & $1/2$ & -4.260 & 10.4 & Yes \\
        144.385268 & 2 & 0 & 2 & $3/2$ & $5/2$ & $5/2$ & 1 & 0 & 1 & $1/2$ & $3/2$ & $3/2$ & -3.664 & 10.4 & Yes \\
        144.390759 & 2 & 0 & 2 & $3/2$ & $1/2$ & $3/2$ & 1 & 0 & 1 & $3/2$ & $1/2$ & $3/2$ & -4.050 & 10.4 & Yes \\
        144.391233 & 2 & 0 & 2 & $3/2$ & $5/2$ & $3/2$ & 1 & 0 & 1 & $1/2$ & $3/2$ & $1/2$ & -3.854 & 10.4 & Yes \\
        144.391233 & 2 & 0 & 2 & $3/2$ & $1/2$ & $1/2$ & 1 & 0 & 1 & $1/2$ & $1/2$ & $3/2$ & -4.245 & 10.4 & Yes \\
        144.392039 & 2 & 0 & 2 & $3/2$ & $3/2$ & $1/2$ & 1 & 0 & 1 & $1/2$ & $1/2$ & $1/2$ & -4.136 & 10.4 & Yes \\
        144.398146 & 2 & 0 & 2 & $3/2$ & $3/2$ & $3/2$ & 1 & 0 & 1 & $1/2$ & $1/2$ & $3/2$ & -4.089 & 10.4 & Yes \\
        144.398681 & 2 & 0 & 2 & $3/2$ & $5/2$ & $3/2$ & 1 & 0 & 1 & $1/2$ & $3/2$ & $3/2$ & -4.274 & 10.4 & Yes \\
        144.400428 & 2 & 0 & 2 & $3/2$ & $3/2$ & $5/2$ & 1 & 0 & 1 & $3/2$ & $3/2$ & $5/2$ & -4.176 & 10.4 & Yes \\
        144.401458 & 2 & 0 & 2 & $3/2$ & $5/2$ & $5/2$ & 1 & 0 & 1 & $1/2$ & $3/2$ & $5/2$ & -4.139 & 10.4 & Yes \\
        144.415423 & 2 & 0 & 2 & $3/2$ & $5/2$ & $7/2$ & 1 & 0 & 1 & $3/2$ & $5/2$ & $7/2$ & -4.142 & 10.4 & Yes \\
        \hline
        \multicolumn{16}{l}{\footnotesize \textbf{Notes.} Only transitions with intensity $T_\text{A}^* > \sigma/3$ are considered and hence presented in this table.}
    \end{tabular}
\end{table*}

\begin{table*}
    \centering
    \caption{\label{tab:H2CN_spectroscopy} Same as Table~\ref{tab:H2NC_spectroscopy} but for \ce{H2CN}. For this molecule, quantum numbers have been obtained from \citet{Yamamoto1992} and CDMS. As in the case of \ce{H2NC}, all of the transitions belong to ortho-\ce{H2CN}.}
    \begin{tabular}{c c c c c c c c c c c c c c c c}
        \hline
        $\nu_\text{rest}$ & $N'$ & $K_\text{a}'$ & $K_\text{c}'$ & $J'$ & $F_1'$ & $F'$ & $N''$ & $K_\text{a}''$ & $K_\text{c}''$ & $J''$ & $F_1''$ & $F''$ & $\log I$ & $E_\text{up}$ & Checked \\
        ($\text{GHz}$) &  &  &  &  &  &  &  &  &  &  &  &  & ($\text{nm}^2 \, \text{MHz}$) & ($\text{K}$)  &  \\
        \hline
        73.324939 & 1 & 0 & 1 & $3/2$ & $3/2$ & $5/2$ & 0 & 0 & 0 & $1/2$ & $3/2$ & $5/2$ & -5.427 & 3.5 & No \\
        73.328925 & 1 & 0 & 1 & $3/2$ & $1/2$ & $3/2$ & 0 & 0 & 0 & $1/2$ & $3/2$ & $3/2$ & -5.412 & 3.5 & No \\
        73.342507 & 1 & 0 & 1 & $3/2$ & $1/2$ & $3/2$ & 0 & 0 & 0 & $1/2$ & $1/2$ & $1/2$ & -4.856 & 3.5 & No \\
        73.345486 & 1 & 0 & 1 & $3/2$ & $3/2$ & $5/2$ & 0 & 0 & 0 & $1/2$ & $3/2$ & $3/2$ & -4.636 & 3.5 & No \\
        73.349203 & 1 & 0 & 1 & $3/2$ & $1/2$ & $1/2$ & 0 & 0 & 0 & $1/2$ & $1/2$ & $3/2$ & -5.081 & 3.5 & No \\
        73.349648 & 1 & 0 & 1 & $3/2$ & $5/2$ & $7/2$ & 0 & 0 & 0 & $1/2$ & $3/2$ & $5/2$ & -4.446 & 3.5 & No \\
        73.355762 & 1 & 0 & 1 & $3/2$ & $7/2$ & $3/2$ & 0 & 0 & 0 & $1/2$ & $3/2$ & $1/2$ & -5.407 & 3.5 & No \\
        73.369366 & 1 & 0 & 1 & $3/2$ & $5/2$ & $3/2$ & 0 & 0 & 0 & $1/2$ & $1/2$ & $3/2$ & -4.863 & 3.5 & Yes \\
        73.392507 & 1 & 0 & 1 & $3/2$ & $3/2$ & $1/2$ & 0 & 0 & 0 & $1/2$ & $3/2$ & $1/2$ & -5.102 & 3.5 & Yes \\
        73.395101 & 1 & 0 & 1 & $3/2$ & $3/2$ & $3/2$ & 0 & 0 & 0 & $1/2$ & $3/2$ & $1/2$ & -4.872 & 3.5 & Yes \\
        73.408673 & 1 & 0 & 1 & $3/2$ & $3/2$ & $3/2$ & 0 & 0 & 0 & $1/2$ & $1/2$ & $3/2$ & -5.439 & 3.5 & No \\
        73.409042 & 1 & 0 & 1 & $3/2$ & $5/2$ & $5/2$ & 0 & 0 & 0 & $1/2$ & $1/2$ & $3/2$ & -4.588 & 3.5 & No \\
        73.444240 & 1 & 0 & 1 & $1/2$ & $3/2$ & $5/2$ & 0 & 0 & 0 & $1/2$ & $3/2$ & $5/2$ & -4.652 & 3.5 & Yes \\
        73.464764 & 1 & 0 & 1 & $1/2$ & $3/2$ & $5/2$ & 0 & 0 & 0 & $1/2$ & $3/2$ & $3/2$ & -5.445 & 3.5 & No \\
        73.465480 & 1 & 0 & 1 & $1/2$ & $3/2$ & $3/2$ & 0 & 0 & 0 & $1/2$ & $3/2$ & $3/2$ & -4.870 & 3.5 & No \\
        73.479075 & 1 & 0 & 1 & $1/2$ & $3/2$ & $3/2$ & 0 & 0 & 0 & $1/2$ & $1/2$ & $1/2$ & -5.423 & 3.5 & No \\
        73.485656 & 1 & 0 & 1 & $1/2$ & $3/2$ & $1/2$ & 0 & 0 & 0 & $1/2$ & $1/2$ & $1/2$ & -5.060 & 3.5 & No \\
        73.505877 & 1 & 0 & 1 & $1/2$ & $1/2$ & $1/2$ & 0 & 0 & 0 & $1/2$ & $3/2$ & $3/2$ & -5.061 & 3.6 & No \\
        73.510462 & 1 & 0 & 1 & $1/2$ & $1/2$ & $3/2$ & 0 & 0 & 0 & $1/2$ & $3/2$ & $5/2$ & -4.761 & 3.6 & No \\
        146.632787 & 2 & 0 & 2 & $5/2$ & $3/2$ & $1/2$ & 1 & 0 & 1 & $3/2$ & $3/2$ & $1/2$ & -5.590 & 10.6 & No \\
        146.640124 & 2 & 0 & 2 & $5/2$ & $5/2$ & $3/2$ & 1 & 0 & 1 & $3/2$ & $5/2$ & $5/2$ & -5.406 & 10.6 & No \\
        146.640480 & 2 & 0 & 2 & $5/2$ & $5/2$ & $3/2$ & 1 & 0 & 1 & $3/2$ & $3/2$ & $3/2$ & -5.499 & 10.6 & No \\
        146.643059 & 2 & 0 & 2 & $5/2$ & $5/2$ & $3/2$ & 1 & 0 & 1 & $3/2$ & $3/2$ & $1/2$ & -5.569 & 10.6 & No \\
        146.649519 & 2 & 0 & 2 & $5/2$ & $5/2$ & $7/2$ & 1 & 0 & 1 & $3/2$ & $5/2$ & $7/2$ & -4.933 & 10.6 & No \\
        146.656268 & 2 & 0 & 2 & $5/2$ & $3/2$ & $5/2$ & 1 & 0 & 1 & $3/2$ & $3/2$ & $5/2$ & -4.915 & 10.6 & No \\
        146.660741 & 2 & 0 & 2 & $5/2$ & $7/2$ & $5/2$ & 1 & 0 & 1 & $3/2$ & $5/2$ & $5/2$ & -4.605 & 10.6 & No \\
        146.661097 & 2 & 0 & 2 & $5/2$ & $7/2$ & $5/2$ & 1 & 0 & 1 & $3/2$ & $3/2$ & $3/2$ & -5.134 & 10.6 & No \\
        146.669522 & 2 & 0 & 2 & $5/2$ & $3/2$ & $1/2$ & 1 & 0 & 1 & $3/2$ & $5/2$ & $3/2$ & -5.257 & 10.6 & No \\
        146.672825 & 2 & 0 & 2 & $5/2$ & $3/2$ & $5/2$ & 1 & 0 & 1 & $3/2$ & $1/2$ & $3/2$ & -3.942 & 10.6 & No \\
        146.674203 & 2 & 0 & 2 & $5/2$ & $5/2$ & $7/2$ & 1 & 0 & 1 & $3/2$ & $3/2$ & $5/2$ & -3.804 & 10.6 & No \\
        146.675065 & 2 & 0 & 2 & $5/2$ & $7/2$ & $9/2$ & 1 & 0 & 1 & $3/2$ & $5/2$ & $7/2$ & -3.675 & 10.6 & No \\
        146.679794 & 2 & 0 & 2 & $5/2$ & $5/2$ & $3/2$ & 1 & 0 & 1 & $3/2$ & $5/2$ & $3/2$ & -4.424 & 10.6 & No \\
        146.689681 & 2 & 0 & 2 & $5/2$ & $3/2$ & $1/2$ & 1 & 0 & 1 & $3/2$ & $1/2$ & $1/2$ & -4.492 & 10.6 & No \\
        146.690331 & 2 & 0 & 2 & $3/2$ & $3/2$ & $5/2$ & 1 & 0 & 1 & $1/2$ & $1/2$ & $3/2$ & -5.206 & 10.6 & No \\
        146.699950 & 2 & 0 & 2 & $5/2$ & $5/2$ & $3/2$ & 1 & 0 & 1 & $3/2$ & $1/2$ & $1/2$ & -4.439 & 10.6 & No \\
        146.700407 & 2 & 0 & 2 & $5/2$ & $7/2$ & $5/2$ & 1 & 0 & 1 & $3/2$ & $5/2$ & $3/2$ & -4.030 & 10.6 & No \\
        146.706337 & 2 & 0 & 2 & $5/2$ & $3/2$ & $3/2$ & 1 & 0 & 1 & $3/2$ & $3/2$ & $3/2$ & -4.766 & 10.6 & No \\
        146.708937 & 2 & 0 & 2 & $5/2$ & $3/2$ & $3/2$ & 1 & 0 & 1 & $3/2$ & $3/2$ & $1/2$ & -4.203 & 10.6 & No \\
        146.712990 & 2 & 0 & 2 & $5/2$ & $5/2$ & $5/2$ & 1 & 0 & 1 & $3/2$ & $5/2$ & $5/2$ & -4.867 & 10.6 & No \\
        146.713348 & 2 & 0 & 2 & $5/2$ & $5/2$ & $5/2$ & 1 & 0 & 1 & $3/2$ & $3/2$ & $3/2$ & -3.972 & 10.6 & No \\
        146.721851 & 2 & 0 & 2 & $5/2$ & $7/2$ & $7/2$ & 1 & 0 & 1 & $3/2$ & $5/2$ & $5/2$ & -3.791 & 10.6 & No \\
        146.728893 & 2 & 0 & 2 & $3/2$ & $1/2$ & $3/2$ & 1 & 0 & 1 & $1/2$ & $1/2$ & $1/2$ & -4.848 & 10.6 & No \\
        146.744745 & 2 & 0 & 2 & $3/2$ & $5/2$ & $5/2$ & 1 & 0 & 1 & $1/2$ & $1/2$ & $3/2$ & -4.456 & 10.6 & No \\
        146.745835 & 2 & 0 & 2 & $3/2$ & $5/2$ & $7/2$ & 1 & 0 & 1 & $1/2$ & $3/2$ & $5/2$ & -3.929 & 10.6 & No \\
        146.748675 & 2 & 0 & 2 & $3/2$ & $5/2$ & $3/2$ & 1 & 0 & 1 & $1/2$ & $1/2$ & $3/2$ & -5.525 & 10.6 & No \\
        146.755839 & 2 & 0 & 2 & $3/2$ & $3/2$ & $5/2$ & 1 & 0 & 1 & $1/2$ & $3/2$ & $3/2$ & -4.088 & 10.6 & No \\
        146.756570 & 2 & 0 & 2 & $3/2$ & $3/2$ & $5/2$ & 1 & 0 & 1 & $1/2$ & $3/2$ & $5/2$ & -5.501 & 10.6 & No \\
        146.762703 & 2 & 0 & 2 & $3/2$ & $1/2$ & $3/2$ & 1 & 0 & 1 & $1/2$ & $3/2$ & $1/2$ & -4.389 & 10.6 & No \\
        146.769300 & 2 & 0 & 2 & $3/2$ & $1/2$ & $3/2$ & 1 & 0 & 1 & $1/2$ & $3/2$ & $3/2$ & -5.191 & 10.6 & No \\
        146.773790 & 2 & 0 & 2 & $3/2$ & $5/2$ & $3/2$ & 1 & 0 & 1 & $1/2$ & $1/2$ & $1/2$ & -4.724 & 10.6 & No \\
        146.785361 & 2 & 0 & 2 & $3/2$ & $3/2$ & $3/2$ & 1 & 0 & 1 & $1/2$ & $1/2$ & $3/2$ & -4.259 & 10.6 & No \\
        146.793082 & 2 & 0 & 2 & $3/2$ & $3/2$ & $1/2$ & 1 & 0 & 1 & $1/2$ & $1/2$ & $1/2$ & -4.471 & 10.6 & No \\
        146.807617 & 2 & 0 & 2 & $3/2$ & $5/2$ & $3/2$ & 1 & 0 & 1 & $1/2$ & $3/2$ & $1/2$ & -4.537 & 10.6 & No \\
        146.810168 & 2 & 0 & 2 & $3/2$ & $1/2$ & $1/2$ & 1 & 0 & 1 & $1/2$ & $1/2$ & $3/2$ & -4.392 & 10.6 & No \\
        146.810268 & 2 & 0 & 2 & $3/2$ & $5/2$ & $5/2$ & 1 & 0 & 1 & $1/2$ & $3/2$ & $3/2$ & -4.813 & 10.6 & No \\
        146.810476 & 2 & 0 & 2 & $3/2$ & $3/2$ & $3/2$ & 1 & 0 & 1 & $1/2$ & $1/2$ & $1/2$ & -5.545 & 10.6 & No \\
        146.810984 & 2 & 0 & 2 & $3/2$ & $5/2$ & $5/2$ & 1 & 0 & 1 & $1/2$ & $3/2$ & $5/2$ & -4.177 & 10.6 & No \\
        146.814198 & 2 & 0 & 2 & $3/2$ & $5/2$ & $3/2$ & 1 & 0 & 1 & $1/2$ & $3/2$ & $3/2$ & -4.544 & 10.6 & No \\
        146.833500 & 2 & 0 & 2 & $3/2$ & $3/2$ & $1/2$ & 1 & 0 & 1 & $1/2$ & $3/2$ & $3/2$ & -5.272 & 10.6 & No \\
        146.840430 & 2 & 0 & 2 & $3/2$ & $5/2$ & $7/2$ & 1 & 0 & 1 & $3/2$ & $5/2$ & $7/2$ & -4.363 & 10.6 & No \\
        146.851599 & 2 & 0 & 2 & $3/2$ & $3/2$ & $3/2$ & 1 & 0 & 1 & $1/2$ & $3/2$ & $5/2$ & -4.675 & 10.6 & No \\
        146.865137 & 2 & 0 & 2 & $3/2$ & $5/2$ & $7/2$ & 1 & 0 & 1 & $3/2$ & $3/2$ & $5/2$ & -5.550 & 10.6 & No \\
        \hline
    \end{tabular}
\end{table*}

\begin{table*}
    \centering
    \contcaption{}
    \begin{tabular}{c c c c c c c c c c c c c c c c}
        \hline
        $\nu_\text{rest}$ & $N'$ & $K_\text{a}'$ & $K_\text{c}'$ & $J'$ & $F_1'$ & $F'$ & $N''$ & $K_\text{a}''$ & $K_\text{c}''$ & $J''$ & $F_1''$ & $F''$ & $\log I$ & $E_\text{up}$ & Checked \\
        ($\text{GHz}$) &  &  &  &  &  &  &  &  &  &  &  &  & ($\text{nm}^2 \, \text{MHz}$) & ($\text{K}$)  &  \\
        \hline
        146.875854 & 2 & 0 & 2 & $3/2$ & $3/2$ & $5/2$ & 1 & 0 & 1 & $3/2$ & $3/2$ & $5/2$ & -4.561 & 10.6 & No \\
        146.892416 & 2 & 0 & 2 & $3/2$ & $3/2$ & $5/2$ & 1 & 0 & 1 & $3/2$ & $1/2$ & $3/2$ & -5.519 & 10.6 & No \\
        146.905561 & 2 & 0 & 2 & $3/2$ & $5/2$ & $5/2$ & 1 & 0 & 1 & $3/2$ & $5/2$ & $7/2$ & -5.121 & 10.6 & No \\
        146.905862 & 2 & 0 & 2 & $3/2$ & $1/2$ & $3/2$ & 1 & 0 & 1 & $3/2$ & $1/2$ & $3/2$ & -4.738 & 10.6 & No \\
        \hline
        \multicolumn{16}{l}{\footnotesize \textbf{Notes.} Only transitions with intensity $T_\text{A}^* > \sigma/3$ are considered and hence presented in this table.}
    \end{tabular}
\end{table*}

\section{\ce{H2CN} excitation temperature}\label{sec:H2CN_tex}

\begin{figure*}
    \centering
    \includegraphics[width=\textwidth]{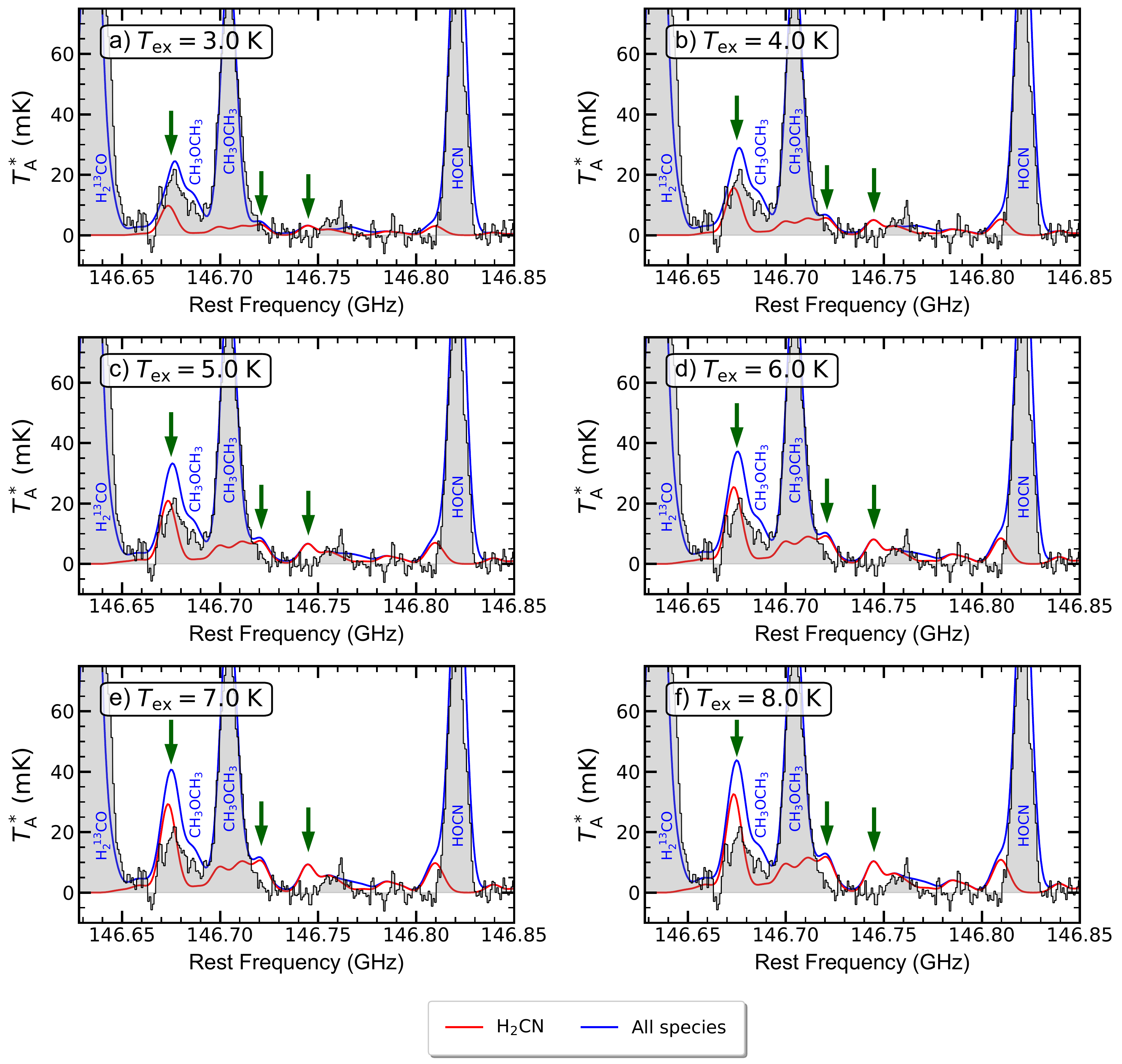}
    \caption{\ce{H2CN} LTE synthetic spectrum for the $2_{02} - 1_{01}$ rotational transition in the range 146$-$147\GHz as a function of $T_\text{ex}$. Black histogram and grey-shaded areas indicate the observed spectrum, while the red and blue lines indicate the emission of \ce{H2CN} and that of all the species already identified in the cloud (including \ce{H2CN}), respectively, whose names are indicated by the blue labels. Green arrows point to the \ce{H2CN} hyperfine transitions that have been used to constrain the $T_\text{ex}$.}
    \label{fig:H2CN_tex} 
\end{figure*}

Fig.~\ref{fig:H2CN_tex} shows the best \ce{H2CN} LTE fit for six different $T_\text{ex}$ ranging 3$-$8\K, focusing on the high-energy $2_{02} - 1_{01}$ transitions. To obtain this fit, we have applied AUTOFIT using the four unblended $1_{01} - 0_{00}$ transitions, while keeping fixed $v_\text{LSR}$ and $\text{FWHM}$ to the values derived for \ce{H2NC} ($v_\text{LSR} = 71.0$\kms and $\text{FWHM} = 18.3$\kms, see Sect.~\ref{sec:H2NC_detection}). It is clear that $T_\text{ex}$ for \ce{H2CN} should be close to 3\K, since higher temperatures clearly overestimate the $2_{02} - 1_{01}$ transitions, as indicated with green arrows in Fig.~\ref{fig:H2CN_tex}. Thus, we have assumed for the \ce{H2CN} fit the same $T_\text{ex}$ as that found for \ce{H2NC} (3.28\K, see Sects.~\ref{sec:H2NC_detection} and \ref{sec:H2CN_detection}). 

\bsp	
\label{lastpage}
\end{document}